\documentclass[twocolumn]{article}

\usepackage[hmarginratio=1:1,top=20mm,left=20mm,right=20mm,columnsep=20pt]{geometry} % Document margins
\usepackage{multicol} % Used for the two-column layout of the document
\usepackage[hang, small,labelfont=bf,up,textfont=it,up]{caption} % Custom captions under/above floats in tables or figures
\usepackage{booktabs} % Horizontal rules in tables
\usepackage{float} % Required for tables and figures in the multi-column environment - they need to be placed in specific locations with the [H] (e.g. \begin{table}[H])
\usepackage[pdftex]{hyperref}
\hypersetup{
pdftitle={Once Upon a Crime: Towards Crime Prediction from Demographics and Mobile Data},
pdfauthor={Andrey Bogomolov, Bruno Lepri, Jacopo Staiano, Nuria Oliver, Fabio Pianesi, Alex Pentland},
pdfkeywords={Crime; Hotspots; Mobile Computing; Demographics},
bookmarksnumbered,
pdfstartview={FitH},
colorlinks
}
\usepackage{graphicx}
\usepackage{float} % Required for tables and figures in the multi-column environment - they need to be placed in specific locations with the [H] (e.g. \begin{table}[H])
\usepackage[ruled]{algorithm2e}

\usepackage{color}

\usepackage{abstract} % Allows abstract customization
\usepackage{titlesec} % Allows customization of titles
\renewcommand\thesection{\Roman{section}} % Roman numerals for the sections
\renewcommand\thesubsection{\Alph{subsection}}
\titleformat{\section}[block]{\bf \scshape\centering}{\thesection.}{1em}{} % Change the look of the section titles
\titleformat{\subsection}[block]{\bf \centering}{\thesubsection.}{1em}{} % Change the look of the section titles
\titleformat{\subsubsection}[block]{\it}{\thesubsubsection.}{1em}{} % Change the look of the section titles
\usepackage{multirow}

\date{}
\begin{document}
%
% --- Author Metadata here ---
% \conferenceinfo{KDD}{2014}
%\CopyrightYear{2007} % Allows default copyright year (20XX) to be over-ridden - IF NEED BE.
%\crdata{0-12345-67-8/90/01}  % Allows default copyright data (0-89791-88-6/97/05) to be over-ridden - IF NEED BE.
% --- End of Author Metadata ---

%\title{Crime Prediction from Aggregated and Anonymized Human Behavioral Data}
\title{\textbf{Once Upon a Crime: Towards Crime Prediction\\from Demographics and Mobile Data}}
\author{Andrey Bogomolov$^\dag$, Bruno Lepri$^\S$,
Jacopo Staiano$^\dag$,\\ Nuria Oliver$^\ddag$, Fabio Pianesi$^\S$, Alex Pentland$^\sharp$ \\
\textit{\small $^\dag$University of Trento (Italy)}\\\textit{\small $^\ddag$Telefonica Research, Barcellona (Spain)}\\\textit{\small $^\S$Fondazione Bruno Kessler, Trento
(Italy)}\\\textit{\small $^\sharp$MIT Media Lab, Cambridge, MA (United States)}}

\twocolumn[
  \begin{@twocolumnfalse}
    \maketitle
\begin{abstract}
In this paper, we present a novel approach to predict crime in a geographic space from multiple data sources, in particular mobile phone and demographic data. 
%We focus on \emph{hotspot} forecasting, \emph{i.e.} predicting which urban areas are likely to become the "scene of a crime".
The main contribution of the proposed approach lies in using aggregated and anonymized human behavioral data derived from mobile network activity to tackle the crime prediction 
problem. 
While previous research efforts have used either background historical knowledge or offenders' profiling, our findings support the hypothesis that aggregated human behavioral data 
captured from the mobile network infrastructure, in combination with basic demographic information, can be used to predict crime. 
%Hence, we propose to combine this type of information with traditional data sources, such as historic socio-economic factors in order to improve the predictive performance of 
% statistical models of crime. 
In our experimental results with real crime data from London we obtain an accuracy of almost 70\% when predicting whether a specific area in the city will be a crime hotspot or 
not. Moreover, we provide a discussion of the implications of our findings for data-driven crime analysis.

%This paper provides the first evidence that next month crime level prediction
%$can be done from aggregated and anonymized mobile network activity in a more
%$accurate way than based on official statistics, collected from
%borough profiles for London Metropolian Area.

%We introduce an original methodology for feature extraction, selection and
%prediction model development in application to crime prediction task, formalized
%as a binary class classification problem. The implications of using this kind of
%mobile network big data for social good are discussed.
\end{abstract}
  \end{@twocolumnfalse}
]

% \category{I.5}{Computing Methodologies}{PATTERN RECOGNITION}[Models, Design
% Methodology, Implementation]
% \category{J.4}{Computer Applications}{SOCIAL AND BEHAVIORAL
% SCIENCES}[Sociology]

% \terms{Algorithms; Experimentation; Measurement; Theory}

% \keywords{crime prediction; urban computing; mobile sensing} % NOT required for Proceedings
% \vspace{1cm}

% \begin{multicols}{2} % Two-column layout throughout the main article text
\section{Introduction}
\label{sec:intro}
Crime, in all its facets, is a well-known social problem affecting the quality
of life and the economic development of a society. Studies have shown that crime tends to be associated with slower economic growth at both the national level \cite{Mehlum2005} and the local level, such as cities and metropolitan areas \cite{Cullen1999}. 
Crime-related information has always attracted the attention
of criminal law and sociology scholars. Dating back to the beginning of the
20th century, studies have focused on the behavioral evolution of
criminals and its relations with specific characteristics of the neighborhoods
in which they grew up, lived, and acted. 

The study of the impact on behavioral
development of factors like exposure to specific peer networks, neighborhood characteristics (\emph{e.g.} presence/absence of recreational/educational facilities)
and poverty indexes, has provided a wealth of knowledge from both individual and
collective standpoints \cite{weinberg1954theories}.
Existing works in the fields of criminology, sociology, psychology and economics tend to
mainly explore relationships between criminal activity and socio-economic
variables such as education \cite{Ehrlich1975}, ethnicity \cite{Braithwaite1989}, income level \cite{Patterson1991}, and unemployment \cite{Freeman1999}.

%Recently, a tendency towards the use of data-driven
%approaches to \emph{crime prediction} has spawned, leveraging the increasing
%availability of data related to crime, places and both collective and individual behaviors.

%Moreover, some researchers proposed to switch the
%popular people-centric paradigm of police practices to a place-centric one
%\cite{Weisburd2008}

%Theoretically, \emph{place-based policing} is based on the \emph{routine
%activities theory} \cite{CohenFelson79} which identifies a crime as the convergence of suitable
%targets (\emph{e.g.} the victims), an absence of capable guardians (\emph{e.g.} the police),
%and the presence of motivated and potential offenders. This all must occur in
%the context of a place or situation and accordingly \emph{place-based policing}
%recognises that topology and street network structure affects the convergence of
%these elements (\cite{brantingham1981environmental} and
%\cite{brantingham1984patterns}).

Several studies in criminology and sociology have provided evidence of
significant concentrations of crime at micro levels of geography, regardless of
the specific unit of analysis defined
\cite{brantingham1999theoretical,weisburd1994defining}.
It is important to note that such clustering of crime in small geographic areas (\emph{e.g.} streets), commonly referred to as \emph{hotspots}, does not necessarily align with trends that are occurring at a larger geographic level, such as
communities. Research has shown, for example, that in what are generally seen as
good parts of town there are often streets with strong crime concentrations, and
in what are often defined as bad neighborhoods, many places are relatively free
of crime \cite{weisburd1994defining}.

In 2008, criminologist David Weisburd proposed to switch the
popular people-centric paradigm of police practices to a place-centric one
\cite{Weisburd2008}, thus focusing on geographical topology and micro-structures
rather than on criminal profiling. 
%In previous studies, the term \emph{crime
%prediction} has mostly been used when estimating the
%evolution of the criminal behavior of an individual in order to make informed decisions on \emph{e.g.} custody, observation, and alert police practices. 
In our paper, \emph{crime prediction} is used in conjunction with a place-centric definition of the problem and with a data-driven approach: 
we specifically investigate the predictive power of aggregated and anonymized human behavioral 
data derived from a multimodal combination of mobile network activity and 
demographic information to determine whether a geographic area is likely to become a \emph{scene of the crime} or not.

As the number of mobile phones actively in use worldwide approaches the 6.8
billion mark\footnote{http://www.itu.int},
they
become a very valuable and unobtrusive source of human behavioral data: 
mobile phones can be seen as sensors of aggregated 
human activity \cite{Dong2011,Laurila2013} and have been used to monitor citizens' mobility patterns and urban interactions \cite{Gonzalez2008,Zheng2014}, to understand individual spending behaviors \cite{Singh2013}, to infer people's traits \cite{Montjoye2013,Staiano2012} and states \cite{Bogomolov2013}, to map and model the spreading of diseases such as malaria \cite{Wesolowski2012} and H1N1 flu \cite{FriasMartinez2011}, and to predict and understand socio-economic indicators of territories \cite{Eagle2010,Soto2011,Capra2014}.
Recently, Zheng \emph{et al.} proposed a multi-source approach, based on human mobility and geographical data, to infer noise pollution \cite{ZhengUbicomp2014} and gas consumption \cite{ShangKDD2014} in large metropolitan areas.

%derived from the devices connections to cell towers, is available to telco
%operators in form of Call Data Records (CDR).
%The information stored in CDRs and available to a mobile network operator
%include the Base Transceiver Stations (BTS) to which callers and recipients of a
%phone call were connected to. Moreover, in the specific case of mobile phones,
%which constantly connect to the closest BTSs with varying signal strength, such
%information allows to identify, by means of triangulation, the approximate path
%followed by an individual. Thus, while the information stored in CDRs serves for
%customers billing, it has the potential of being extremely useful for other
%purposes.

In this paper, we use human behavioral data derived from a combination of mobile network activity and demography, together with open
data related to crime events to predict crime \emph{hotspots} in specific neighborhoods of a European metropolis: London. 
The main contributions of this work are:
\begin{enumerate}
\item The use of human behavioral data derived from 
anonymized and aggregated mobile network activity, combined with demographics, to predict crime hotspots in a European metropolis.
\item A comprehensive analysis of the predictive power of the proposed model and a comparison with a state-of-the-art approach based on official statistics.
\item A discussion of the theoretical and practical implications of our proposed approach.
\end{enumerate}

%\fabio{the  two first points seem to be just one. I would drop the first}
%For the task at hand, 
%the availability of comprehensive
%historical crime data provides a solid foundation to build statistical models able to
%predict the evolution of specific types of crime in small geographic areas. In
%our experiments, 
%In particular, we train and validate our models with crime-related information for 
%two months (precisely, December 2012 and January 2013)
%: hence, our
%models do not incorporate historical hotspot knowledge, which allows us to
%independently assess the effectiveness of mobile phone-based data for hotspot
%forecasting. 

%bruno{enforce the novelty of our approach}

This paper is structured as follows: section \ref{sec:relworks} describes   relevant previous work in the area of data-driven crime prediction; 
the data used for our experiments are described in section \ref{sec:Dataset}; 
the definition of the research problem
tackled in this work and detailed information on the methodology adopted is
provided in \ref{sec:Methodology}; finally, we report our experimental results and
provide a discussion thereof in sections \ref{sec:ExperimentalResults} and
\ref{sec:Discussion}, respectively.

\section{Related Work}
\label{sec:relworks}
Researchers have devoted attention to the study of criminal
behavior dynamics both from a people- and place- centric perspective.
The people-centric perspective has mostly been used for
individual or collective criminal profiling. Wang \emph{et al.} \cite{Wang2013}
proposed \emph{Series Finder}, a machine learning approach to the problem
of detecting specific patterns in crimes that are committed by the same offender
or group of offenders. %The goal in this case is to provide au- tomatic tools for crime analysts to determine which crime(s) may have been committed by the same individual(s). 
%The proposed approach has had promising results on a decade's worth of crime pattern
%data collected by the Crime Analysis Unit of the Cambridge Police Department.
In \cite{Short2008}, it is proposed a biased random walk model built upon
empirical knowledge of criminal offenders behavior along with spatio-temporal
crime information to take into account repeating patterns in historical
crime data. Furthermore, Ratcliffe~\cite{ratcliffe2006temporal} investigated the spatio-temporal constraints underlying offenders' criminal behavior.

An example of a place-centric perspective is 
crime hotspot detection and analysis and the
consequent derivation of useful insights.
A novel application of quantitative tools from mathematics, physics and signal
processing has been proposed by Toole \emph{et al.} \cite{Toole2011} to analyse
spatial and temporal patterns in criminal offense records. The analyses they
conducted on a dataset containing crime information from 1991 to 1999 for the
city of Philadelphia, US, indicated the existence of multi-scale complex
relationships both in space and time.
Using demographic information statistics at community (town) level, 
Buczak and Gifford~\cite{buczak2010fuzzy} applied fuzzy association rule mining in order to derive a finite (and consistent among US states) set of rules to be applied by crime analysts. Other common models are the ones proposed by Eck \emph{et al.} \cite{Eck2005} and by Chainey \emph{et al.} \cite{Chainey2008} that rely on kernel density estimation from the criminal history record of a geographical area. Similarly, Mohler \emph{et al.} \cite{Mohler2011} applied the self-exciting point process model (previously developed for earthquake prediction) as a model of crime. The major problem of all these approaches is that they relies on the prior occurrence of crimes in a particular area and thus cannot generalize to previously unobserved areas. 

More recently, the proliferation of social media has sparked interest in using
this kind of data to predict a variety of variables, including electoral outcomes
\cite{tumasjan2010predicting} and market trends \cite{bollen2011twitter}. In
this line, Wang \emph{et al.} \cite{wang2012automatic} proposed the
usage of social media to predict criminal incidents. Their approach relies on
a semantic analysis of tweets using natural language processing along with
spatio-temporal information derived from neighborhood demographic data and the 
tweets metadata. 

%Evaluation results demonstrate the model's ability to forecast
%hit-and-run crimes using only the information contained in the training set of
%tweets, including their content.

%As the deployment of personal communication devices increased dramatically, researchers took the opportunity to use them as sensors for behavioral studies. Thus, another line of works relevant for our paper is the bulk of studies employing mobile data to analyse human behavior. Such works have shown, for instance, that mobile data metrics are predictive of behavioral determinants such as personality~\cite{de2011towards,yves2013pers,staiano2012friends}.

%Call Data Records (CDRs) have been extensively studied for a broad range of purposes, from understanding human mobility [6,11,7,26] to land use identification
%and urban planning [4,24,20]. \jacopo{FIX BIBTEX}
%Various ways of characterising geography based on the traffic of mobile phones and their users' trajectories have been examined. 
%Specific to understanding the relation between CDRs and socioeconomic factors

%Furthermore, the relation between CDRs and socio-economic factors has recently received significant attention from both industry and the research community. For instance, the data made available during open challenges such as D4D~\cite{blondel2012data} has allowed researchers to model the epidemic spreading of diseases using mobility patterns derived from CDRs~\cite{lima2013exploiting}. 

In this paper, we tackle the crime hotspot forecasting problem by leveraging mobile network activity as a source of 
human behavioral data.
Our work hence complements the above-mentioned research efforts and contributes to advance the state-of-the-art in quantitative criminal studies.

\section{Datasets}
\label{sec:Dataset}

% The big data for social good datasets we exploited have been provided during a
% public competition -- the datathon organized by Telef\'{o}nica, The Open Data
% Institute and the MIT during the Campus Party Europe 2013, Europe's largest
% technology festival at the O2 Arena in London during 2-7 September 2013, hosting
% more than 10000 ``campuseros'' -- some of the brightest young minds in tech from
% across Europe. The participants were provided access to a large scale anonymized
% and aggregated data from Telef\'{o}nica mobile network, including:
% \begin{itemize}
%   \item footfall counts for London Metropolitan Area over the course of 3 weeks,
% gender and age group splits for each of the locations within the London
% Metropolitan area on an hourly basis, inference data of crowd split by home,
% work and visitor based on the O2 network activity;
%   \item geo-localised Open Data sets including criminal cases reported,
%   residential property sales, transportation, London borough profiles datasets about
% homelessness, households, housing market, local government finance and societal
% wellbeing;
%   \item non-localised hash-tagged Twitter data sets and
%   \item Telef\'{o}nica O2 network grid output centroids. 
% \end{itemize}

The datasets  we exploit in this paper were provided during a
public competition - the Datathon for Social Good - organized by Telef\'{o}nica Digital, The Open Data Institute and 
MIT during the Campus Party Europe 2013 at the O2 Arena in London in September 2013.
%The Campus Party Europe is Europe's largest
%technology festival hosting
%more than 10000 ``campuseros'' including some of the brightest 
%young minds in tech from all across Europe.
 
Participants 
 %in the Datathon 
were provided access to the following data, among others: 
% all of it for the same time period 
% (3 weeks: from Dec 9th to 15th, 2012 and from Dec 23rd, 2012 to Jan 5th, 2013)
%and for the London Metropolitan area:
% \jacopo{FIXING: not true! geo-localised open data is not for the 3-weeks period!%.}

\begin{itemize}
  \item Anonymized and aggregated human behavioral data computed from mobile network activity in the London Metropolitan Area. We shall 
  refer to this data as the Smartsteps dataset, because it was derived from Telefonica Digital's Smartsteps product\footnote{http://dynamicinsights.telefonica.com/488/smart-steps}. 
  A sample visualization of the Smartsteps product can be seen in Figure \ref{fig:smartsteps};
  
  %The London Metropolitan Area is divided in a grid of 
  %200m by 200m square cells whose centroids are provided. 
  %The \emph{footfall} --estimated number of people within each cell--
  %is made available on an hourly basis. This estimation is derived 
  %from the Telefonica O2 UK mobile network activity. 
  %In addition, gender, age and home/work/visitor group splits
  %are provided hourly and for each of the cell locations. 
  %That is, for each cell in the grid and for each hour, the dataset contains
  %an estimation of how many people are the cell, the percentage of 
  %these people who are at home, at work or just visiting the cell and 
  %their gender and age splits in the following brackets: 0-20 years, 21-30 years, 31-40 years, etc...
  %We shall refer to this dataset as the \emph{SmartSteps dataset} and to the 
  %grid as the SmartSteps grid, because it 
  %was derived from Telefonica Digital's SmartSteps product\footnote{http://dynamicinsights.telefonica.com/488/smart-steps}.
  
  %\item Non-localised hash-tagged Twitter data, which 
  %includes the number of tweets per hour and a sample of 2800 tweets both with the London hashtag;
  
  \item Geo-localised Open Data, a collection of openly available datasets with varying temporal granularity. This includes reported criminal cases,
  residential property sales, transportation, weather and London borough profiles related to
homelessness, households, housing market, local government finance and societal
wellbeing (a total of 68 metrics).
% AB
  %\item Telef\'{o}nica O2 network grid output centroids. 
  %\{isn't the grid already described above?}
\end{itemize}

%Within the Open Data dataset there was the Criminal Cases Dataset, which we used as ground truth for our experiments. It provides crime information aggregated at monthly level for December 2012 and January 2013. The Smartsteps and Twitter data referred to a shorter period of time -- namely, 3 weeks: from December 9th to 15th, 2012 and from December 23rd, 2012 to January 5th, 2013.

We turn now to describing  the specific datasets that we used to predict crime hotspots.

%In the experiments discussed in this paper, 
%we have used the SmartSteps, the profiles of London's boroughs, 
%the reported criminal cases and the Twitter datasets, together with the 
%SmartSteps grid.

%The weather data provided in the competition was not used for
%prediction modelling because prediction and training time intervals were very short and
%the weather was similar to learn it's influence on the crime level.
%Also the spatial points provided did not cover most of the London Metropolian Area - the
%territory we were working on for data analysis and prediction.

%is a spatial dataset of telecom output areas,
%such as location id along with latitude, longitude and area covered in square
%km. The dataset and it's description did not provide any clue about the shape of
%the area covered, the limitations of signal propagation due to the landscape,
%high buildings or interference. We addressed this challenge as described in
%the methodology.
% 
%London borough profiles is an official open dataset about homelessness,
%households, housing market, local government finance and societal wellbeing.
%For the purpose of the result discussed in this paper we used
%Telef\'{o}nica O2 network grid, London borough profiles, twitter, smartsteps and
%reported crime cases data.

\subsection{Smartsteps Dataset}
The Smartsteps dataset consists of a geographic division of the 
London Metropolitan Area into cells whose precise location (lat,long) and
surface area is provided. Note that the actual shape of the cell was not provided. 
%to Datathon participants and therefore it is unknown. 
In total, there were 124119 cells. We shall refer to these cells as the Smartsteps cells.
%\nuria{please add the number of cells}

For each of the Smartsteps cells, a variety of demographic 
variables were provided, computed every hour for a 3-week period, from December 9th to 15th, 2012 and from December 23rd, 2012 to January 5th, 2013.

In particular: 
%The London Metropolitan Area is divided in a grid of 
%  200m by 200m square cells whose centroids are provided. 
%  We shall refer to this grid as the SmartSteps grid.
  
  (1) \emph{Footfall}, or the estimated number of people within each cell.
  This estimation is derived 
  from the 
  %Telefonica O2 UK 
  mobile network activity by aggregating every hour 
  the total number of unique phonecalls in each cell tower, mapping the 
  cell tower coverage areas to the Smartsteps cells, and extrapolating 
  to the general population --by taking into account the market share 
  of the network in each cell location; and 
  
  (2) an estimation of gender, age and home/work/visitor group splits. 
  
  That is, for each Smartsteps cell and for each hour, the dataset contains an estimation of how many people are in the cell, the percentage of 
  these people who are at home, at work or just visiting the cell and 
  their gender and age splits in the following brackets: 0-20 years, 21-30 years, 31-40 years, etc..., as shown in Table \ref{tab:smartsteps}. 
 
 %\nuria{would it make sense to show a picture of the grid with the estimations so reviewers see what we are talking about?}
 %\fabio{yes}
 
%\bruno{, maybe it could make sense make reference to previous applications of smartsteps if any ... e.g. marketing?} 

%total number of people in an area, number of people at home,number of people at work, number of people visiting an area, number of men, number of women, number of people with age between 0 and 20 years, number of people with age between 21 and 30 years, number of people with age between 31 and 40 years, number of people with age between 41 and 50 years, number of people with age between 51 and 60 years and number of people with age more than 60 years

\begin{table}[H]
\caption{SmartSteps data provided by the challenge organizers. All data refer to 1-hour intervals and to each Smartsteps cell.} 
\label{tab:smartsteps}
\footnotesize
\centering
\begin{tabular}{  c | r }
%\hline
\textbf{Type} & \textbf{Data}\\
\hline

\multirow{4}{*}{Origin-based}
      & total \# people\\
      & \# residents \\
      & \# workers \\
      & \# visitors \\
\hline

\multirow{2}{*}{Gender-based}
      & \# males\\
      & \# females \\
\hline

\multirow{6}{*}{Age-based}
      & \# people aged up to 20\\
      & \# people aged 21 to 30 \\
      & \# people aged 31 to 40 \\
      & \# people aged 41 to 50 \\
      & \# people aged 51 to 60 \\
      & \# people aged over 60 \\
%\hline
\end{tabular}
\end{table}

\begin{figure*}[!ht]
\centering
\includegraphics[scale=0.38,trim=0cm 6cm 0cm 0cm, clip=true]{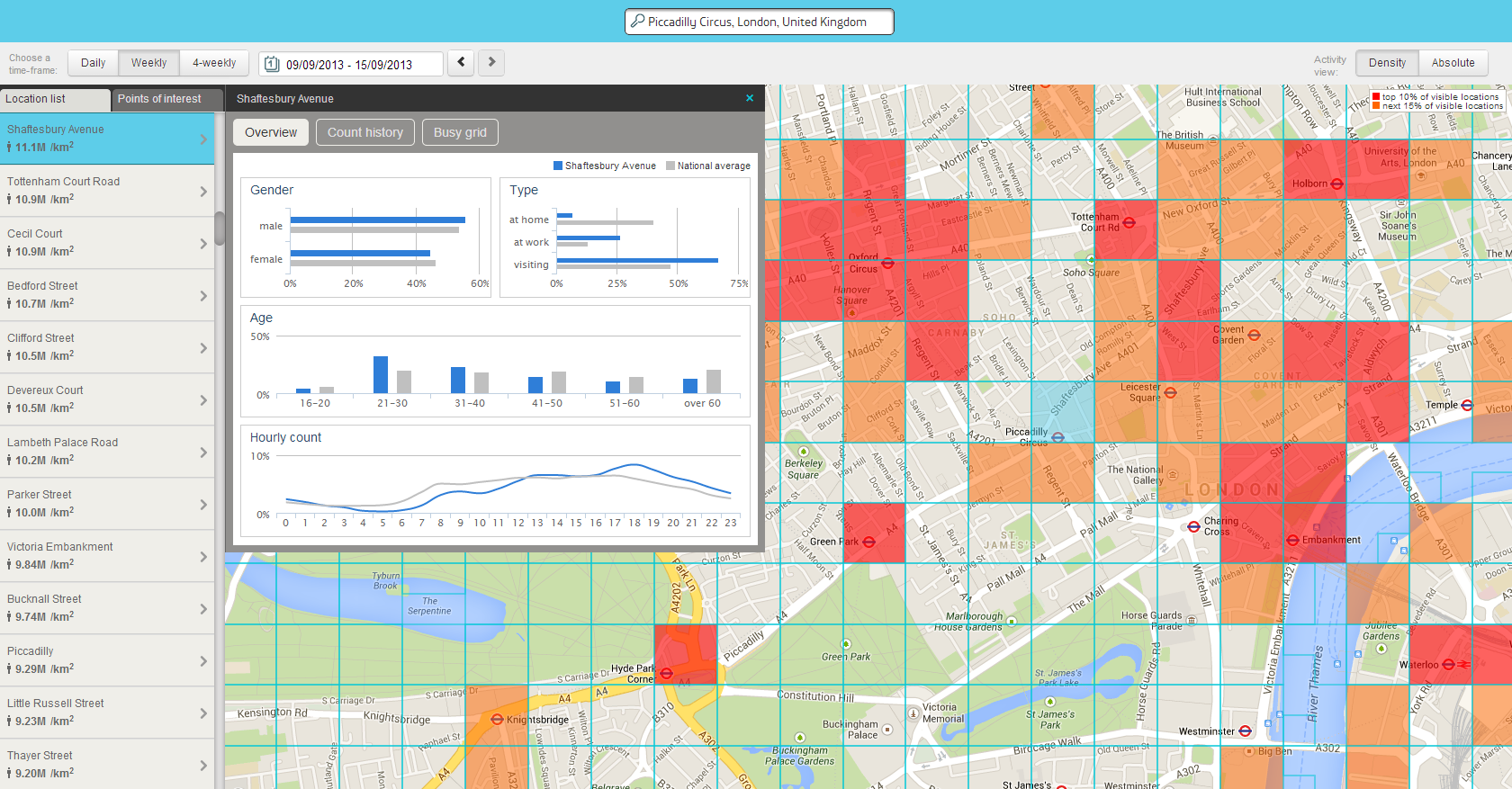}
\caption{Sample visualization of the high-level information provided by the Smartsteps platform. By combining aggregated and anonymized demographics and mobility data, 
fine-grained 
spatio-temporal dynamics can be exploited to derive valuable insights for the scenario of interest.}
\label{fig:smartsteps}
\end{figure*}

Figure~\ref{fig:smartsteps} shows a sample visualization of the information made available from the SmartSteps platform.

%\{isn't the telefonica grid the 200m x 200m grid described in the smartsteps data bulletpoint. 
%They provided the footfall for each cell of this grid, right???}
%The Telef\'{o}nica O2 network grid is a spatial dataset of telecom output areas,
%such as location id along with latitude, longitude and area covered in square
%km. The dataset and its description did not provide any clue about the shape of
%the area covered, the limitations of signal propagation due to the landscape,
%high buildings or interference. We addressed this challenge as described in
%the methodology.

\subsection{Criminal Cases Dataset}

The criminal cases dataset includes the geo-location of all reported crimes in the UK but does not specify their exact date, just the month and year. The data provided in the public competition included the criminal cases for December 2012 and January 2013.

In detail, the crime dataset includes: 
the crime ID, the month and year when the crime was committed, 
its location (longitude, latitude, and address where the crime took place), 
the police department involved, the \emph{lower layer super output area} code, 
the \emph{lower layer super output area} name and its type out of 11 possible types
(\emph{e.g.} anti-social behavior, burglary, violent crime, shoplifting, etc.). %In total 11 crime types are reported.
% \fabio{how many different types are there?} => 
% [+] crime types number added to the latex, AB

% \{what is lower layer super output??}
%Crimes reported for the UK, include geolocation, but doesn't specify date of
%crime (just month and year). The crime data included: crime ID, month when
%the crime was commited, police department, longitude, latitude, address 
%where the crime happend, lower layer super output area code, lower layer super
%output area name, crime type and last outcome type. 
Lower Layer Super Output Areas (LSOAs) are small geographical areas defined by the United Kingdom Office for National Statistics following the 2001 census. 
%and describe spatial neighborhood
%statistics in the United Kingdom. 
Their aim here is to define areas, based on population levels, whose boundaries would not change over time. 
LSOAs are the smallest type of output areas used
for official statistics, have a mean population of 1,500 and a minimum  population threshold of 1,000.
%an area ranging from Xm by Xm in urban areas to Ym by Ym in rural areas. 
%\jacopo{LSOAs are defined by population density, thus vary in size. I couldnt find information on LSOAs size in London.}
%\{add details of geographical coverage}

%Super Output Areas (SOAs) are a geography for the collection and publication of small area statistics, which represent spatial neighbourhoods. They are a statistical geography and their key aspects are stability and uniformity of size. SOAs avoid the problems caused by the inconsistent and unstable population geography. They are considered to be better for statistical comparison as they are of much more consistent size and each layer has a specified minimum population to avoid the risk of data disclosures. SOAs are not subject to frequent boundary changes, and therefore are more suitable for comparison over time. In addition they build on the existing availability of data for output areas (OAs).
%The SOA layers form a hierarchy based on aggregations of Output Areas. Lower Layer Super Output Areas (LSOAs) were first built using 2001 Census data from groups of Output Areas and have been updated following the 2011 Census.  Middle Layer Super Output Areas (MSOAs) are larger areas created by zone-design software using census data from groups of LSOAs. They fit within local authority boundaries.  Upper Layer Super Output Area (USOAs) were not used in the datasets analyzed.
%There are 181,408 OAs and 34,753 LSOAs in England and Wales. Only the LSOAs that correspond to London's Metropolian area were used in the experiments reported in this paper.

The ground truth crime data used in our experiments corresponds to the crimes reported in January 2013 for each of the Smartsteps cells.
% \fabio{I still do not understand what is the relevance of the LSOAs in our case and what is their relationships to the grain-level of the data we are using}
% \nuria{should it not be December 2012?!!!!!!}

\subsection{London Borough Profiles Dataset}

The London borough profiles dataset is an official open dataset containing 68 different metrics about the population of a particular geographic area. The spatial granularity of the bourough profiles data is at the LSOA level.
% \fabio{please, add details of the geographic granularity of this dataset. is it LSOA level??}

In particular, the information in this dataset includes statistics about the population, households (census), demographics, migrant population, ethnicity, language, employment, NEET (Not in Education, Employment, and Training) people, earnings, volunteering, jobs density, business survival, house prices, new homes, greenspace, recycling, carbon emissions, cars, indices of multiple deprivation, children in out-of-work families, life expectancy, teenage conceptions, happiness levels, political control (\emph{e.g.} proportion of seats won by Labour, LibDem and Conservatives), and election turnout.

\section{Methodology}
\label{sec:Methodology}
We cast the problem of crime hotspot forecasting as a binary classification task. For each Smartsteps cell, we predict whether that particular 
cell will be a crime hotspot 
or not in the next month. In this section we provide details of the experimental setup that we followed.

\subsection{Data Preprocessing and Feature Extraction}
Starting from the bulk of data described in Section \ref{sec:Dataset}, we performed the following  
preprocessing steps.

\subsubsection{Referencing all geo-tagged data to the Smartsteps cells}
As the SmartSteps cell IDs, the borough profiles and the crime event locations
are not spatially linked in the provided datasets, we first 
geo-referenced each crime event by identifying the
Smartsteps cell which is the closest to the location of the crime. 
We carried out a similar process for the borough profile dataset. 
As a result, each crime event and the borough profile information were  linked to one of the Smartsteps cells.

\subsubsection{Smartsteps features}

\textit{Diversity} and \textit{regularity} have been shown to be important in the characterization of different facets of human behavior and, in particular, the concept of entropy has been applied to assess the predictability of mobility \cite{Song2010} and spending patterns \cite{Krumme2013, Singh2013}, the socio-economic characteristics of places and cities \cite{Eagle2010} and some individual traits such as personality \cite{Montjoye2013}.
Hence, for each Smartsteps 
variable (see Table~\ref{tab:smartsteps}) we computed the mathematical functions which characterize the distributions and information theoretic properties of
such variables, \emph{e.g.} mean, median, standard deviation, min and max values and Shannon entropy.
% \fabio{what are the other functions?}

In order to be able to also account for temporal relationships inside the Smartsteps data, the same computations as above were repeated 
on sliding windows of variable length (1-hour, 4-hour and 1 day), producing  
 \textit{second-order features} that help reduce computational complexity
and the feature space itself, while preserving useful data properties.

\subsubsection{London borough profile features}
% \bruno{here we need also a description of these features ... they are used in the prediction so they need to be described}
No data preprocessing was needed for the London borough profiles. Hence, we used the original 68 London borough profile features.

\subsubsection{Crime hotspots ground-truth data}

The distribution of the criminal cases data is reported in Table \ref{tab:crimesummary}. 

%\{which is the second month in a row? january? february?}
% 
% \{add information about amount of cells with LOW and HIGH levels of crime}

\begin{table}[H]
\caption{Number of crime hotspots in January.} 
\label{tab:crimesummary} 
\centering
\begin{tabular}{|c|c|c|c|c|c|}
\hline
Min.&Q1&\textbf{Median}&Mean&Q3&Max.\\
\hline
1&2&\textbf{5}&8.2&10&289\\
\hline
\end{tabular}
\end{table}

%\begin{table}[ht]
%\caption{Number of crime hotspots in the month of January} 
%\label{tab:crimesummary} 
%\centering
%\begin{tabular}{r|r}
%  \hline
% & CRIME SUMMARY\\ 
%  \hline
%Min.   &   1.000   \\ 
% 1st Qu. &   2.000   \\ 
% \bf{Median} &   \bf{5.000}   %\\ 
% Mean   &   8.196   \\ 
% 3rd Qu. &  10.000   \\ 
% Max.   & 289.000   \\ 
%   \hline
%\end{tabular}
%\end{table}
% \fabio{so what? a few words by way of comment? is this relevant to our work?}

Given the high skewness of the distribution (skewness = 5.88, kurtosis = 72.5, mean = 8.2, median = 5; see Table \ref{tab:crimesummary}) and based on previous research on urban crime patterns \cite{Boggs1965}, we split the criminal dataset with respect to its median into two classes: 
a \emph{low} crime (class '0') when the number of crimes 
in the given cell was less or equal to
the median, and a \emph{high} crime (class '1') when the number of crimes in a given cell was larger than the median.
% \fabio{what about the cases that fall on the median? are they in the low of in the high class?}

%The usage of the median rather than the mean is justified by the high skeweness of the crime events distribution
%(skewness = 5.88, kurtosis = 72.5, mean = 8.2, median = 5; 
%see Table \ref{tab:crimesummary}).

Following the empirical distribution, the two resulting classes are approximately balanced (53.15\% for the \emph{high} crime class).

%\fabio{why isn't it 50\%-50\%?}
%\nuria{what if the number is exactly the median?}

\subsection{Feature Selection}

We randomly split all data into training (80\% of data) 
and testing (20\% of data) sets.
In order to accelerate the convergence of the models, we \textit{normalized}
each dimension of the feature vector \cite{Box:1989868}. 

As an initial step, we carried out a \textit{Pearson correlation analysis} to visualize and better understand the relationship 
between variables in the feature space. We found quite a large subset of features with strong mutual correlations and another subset of uncorrelated features. 
There was room, therefore, for feature space reduction. We excluded using  \textit{principal component analysis} (PCA) because the transformation it is based on produces new variables that are difficult to interpret in terms of the original ones, which complicates the interpretation of the results.

We turned to a pipelined \textit{variable selection} approach, based on \textit{feature ranking} and \textit{feature subset selection}, which was perfomed using only data from the training set. 
% feature ranking, each feature is ranked using a metric based on classification performance for a given outcome. 
%Feature subset selection, then, can be implemented through wrappers, filters or embedded methods \cite{janecek2008}. 
%\textit{Wrappers} rely on the performance of a black box algorithm to evaluate the quality of a set of features. 
%Wrapper methods search through the space of feature subsets and calculate the estimated metric of a single learning algorithm for each feature following "leave-one-out" strategy.
%\textit{Filters} are classifier agnostic feature selection methods, which are independent of the later applied machine learning algorithm.
%In \textit{Embedded methods} the feature subset selection and the learning method are interleaved.
%\nuria{why are we describing these possible ways to do feature selection if we are not using them?}

%After a first attempt at feature selection using Adaboost \cite{bartlett2007adaboost}, which turned to be computationally too expensive, we used a
%decision tree classifier based on the Breiman's Random Forest (RF) algorithm
%with a \textit{random selection of features} to split each node, an approach that is known to be as consistent as Adaboost and more robust with respect to noise 
%\cite{bartlett2007adaboost}. 

The metric used for feature ranking was the mean decrease in the \textit{Gini coefficient of inequality}. 
%\nuria{are these percentages for OUR data or taken directly from the paper by Singh?}
This choice was motivated because it outperformed 
other metrics, such as mutual information, information gain and chi-square statistic with an average improvement of approximately 28.5\%, 19\% and 9.2\% respectively \cite{singh2010feature}.  
The Gini coefficient ranges between $0$, expressing perfect equality (all dimensions have the same predictive power) and $1$, expressing
maximal inequality in predictive power. The feature with maximum mean 
decrease in Gini coefficient is expected to have the maximum influence 
in minimizing the out-of-the-bag error. It is known in the literature that 
minimizing the out-of-the-bag error results in maximizing common performance metrics used to evaluate models (\emph{e.g.} accuracy, F1, AUC, etc...) \cite{tuv2009feature}. 

In the subsequent text and tables this metric is presented as a percentage.

The feature selection process produced a reduced subset of 68 features (from an initial pool of about 6000 features), with a reduction in dimensionality of about 90 times with 
respect to the full feature space. The top 20 features selected by the model are included in Table~\ref{tab:top30features}.
%Interestingly, no features from the Twitter dataset were selected to be included in the final 68-dimensional feature vector, probably because of the coarse spatial granularity of the tweets (neighborhood level). 

% \fabio{there are many ways to perform feature selection. whatever we adopt, it must be clear, and this all section is misterious.}
% 
%The features extracted from Twitter data did not go through the feature selection process, probably because of low spatial granularity of the tweets.
% \bruno{Andrey, write one or two sentence that could explain this result}
%after the feature selection process.

\subsection{Model Building}
We trained a variety of classifiers on the training data following 5-fold cross validation strategy: 
logistic regression, support vector machines, neural networks, decision trees, and different implementations of ensembles of tree classifiers with different parameters.

%Logistic regression model, {\em support vector machines
%model}\cite{Vapnik:1995:NSL:211359} with linear and Gaussian radial basis
%\cite{Buhmann:2003:RBF:945834} kernels and multi-layer perceptron {\em neural
%network} did not provide good classification results misclassifying one of the
%classes.

The decision tree classifier based on the Breiman's Random Forest (RF) algorithm
yielded the best performance when compared to all other classifiers. Hence, 
we report the performance results only for the the best model, based on this algorithm.

%The final decision boundary for our classification task is as follows. 

%Given
%an ensemble of tree classifiers
%$h_1(\vec{x}),h_2(\vec{x}),...,h_K(\vec{x})$ and if the training set% is
%drawn at random from the empirical distribution of the random vector $Y,
%\vec{X}$ the margin function is defined as:
%\begin{equation}
%\begin{split}
%mg(\mathbf{X},Y) = avg_k{I(h_k(\mathbf{X})=Y)} - \\
% \operatorname{max}_{j!=Y}avg_kI(h_k(\mathbf{X})=j),
%\end{split}
%\end{equation}
%where $I(\cdot)$ is the characteristic function. The margin function measures
%the distance between the average votes at $(\vec{X},Y)$ for the right class
%and the average vote for any other class. For this model the generalization
%error function is:
%\begin{equation}
%PE^* = P_{\mathbf{X},Y}(mg(\mathbf{X},Y)<0),
%\end{equation}
%where $P_{\vec{X},Y}$ is the probability over $\langle\vec{X},Y\rangle$
%space.
%For any event $A\subset\Omega$ of the feature space the characteristic function
%$I(\cdot)$ of $A$ is:
%\begin{equation}
%I_A(\vec{x}) = \begin{Bmatrix} 1 & {\iff (\vec{x} \subset A)} \\ 0 & otherwise
%\end{Bmatrix} \begin{Bmatrix} 1 & {\iff \exists \vec{x}} \\ 0 & otherwise
%\end{Bmatrix}
%\end{equation}

We took advantage of the well-known performance improvements that are obtained by  growing an ensemble of trees and voting for the most frequent
class. Random vectors were generated before the growth of each tree in the ensemble, 
and a random selection without replacement was performed  \cite{Bagging1996}.

%We used a random selection of features to split each node and voting \cite{breiman1999random}, which is a \textit{computationally efficient} algorithm with 
%outstanding properties exploited during competitions by machine learning community.
% \fabio{this sentence is incomplete!}
The \textit{consistency} of the random forest algorithm has been proven and the algorithm adapts to sparsity in the sense that the rate of \textit{convergence} depends only on the number of strong features and not on the number of noisy or less relevant ones \cite{biau2012analysis}.

% In order to enhance the generalization power of the model, we performed bootstrapping 
% \fabio{how? on what? how does bootstrapping here relates to the sampling with replacement discussed below?}
% \cite{efron93bootstrap}.

% Finally, note that we carry out structural risk minimization -- as opposed to empirical risk 
% minimization -- by sampling the data with replacement 
% for each fold during the learning process instead of working with a regularization penalty. 
% We adopted this strategy in order to prevent data overfitting for our complex feature space.
% %which includes second- and higher-order variables.
% \fabio{nobody has talked so far about second and higher order variables features!! where are they?}

%in such a way that they mimic the empirical
%distributions.  
%and working not with a regularization penalty, but 

% \{not sure we need the algorithm below}
% \TODO{Feature space description}
% The features were extracted from
% second order features
% \subsection{Feature Extraction}
% Hilbert space
% {\ldots}
% \subsection{Model Search}
% {\ldots}

\section{Experimental Results}
\label{sec:ExperimentalResults}
In this section we report the experimental results obtained by the Random Forest trained on different
subsets of the selected features and always on the test set, which was not used during the
training phase in any way.

The performance metrics used to evaluate our approach are: accuracy, F1, and AUC score. As can be seen on Table \ref{tab:MetricsComparison}, the model achieves almost 70\% accuracy when predicting whether 
a particular Smartsteps cell will be a crime hotspot in the 
following month or not. 
Table \ref{tab:MetricsComparison} includes all performance metrics obtained by our model. 

\begin{figure*}[!ht]
\centering
\begin{minipage}[b]{.5\textwidth}
\centering
\includegraphics[scale=0.5]{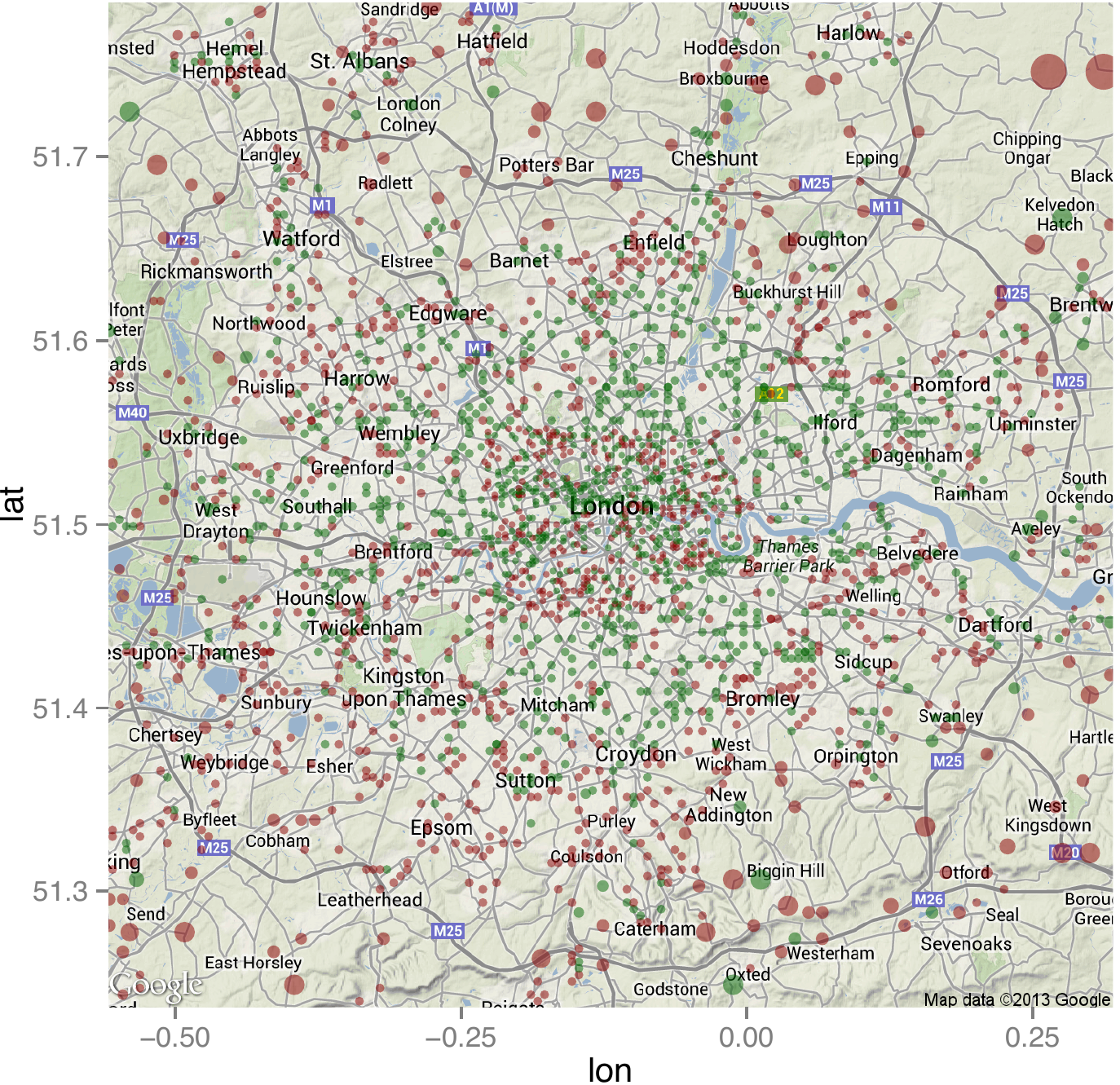}
\caption{Ground Truth of Crime Hotspots}
\label{fig:crimeGroundTruth}
\end{minipage}%
\hfill
\begin{minipage}[b]{.5\textwidth}
\centering
\includegraphics[scale=0.5]{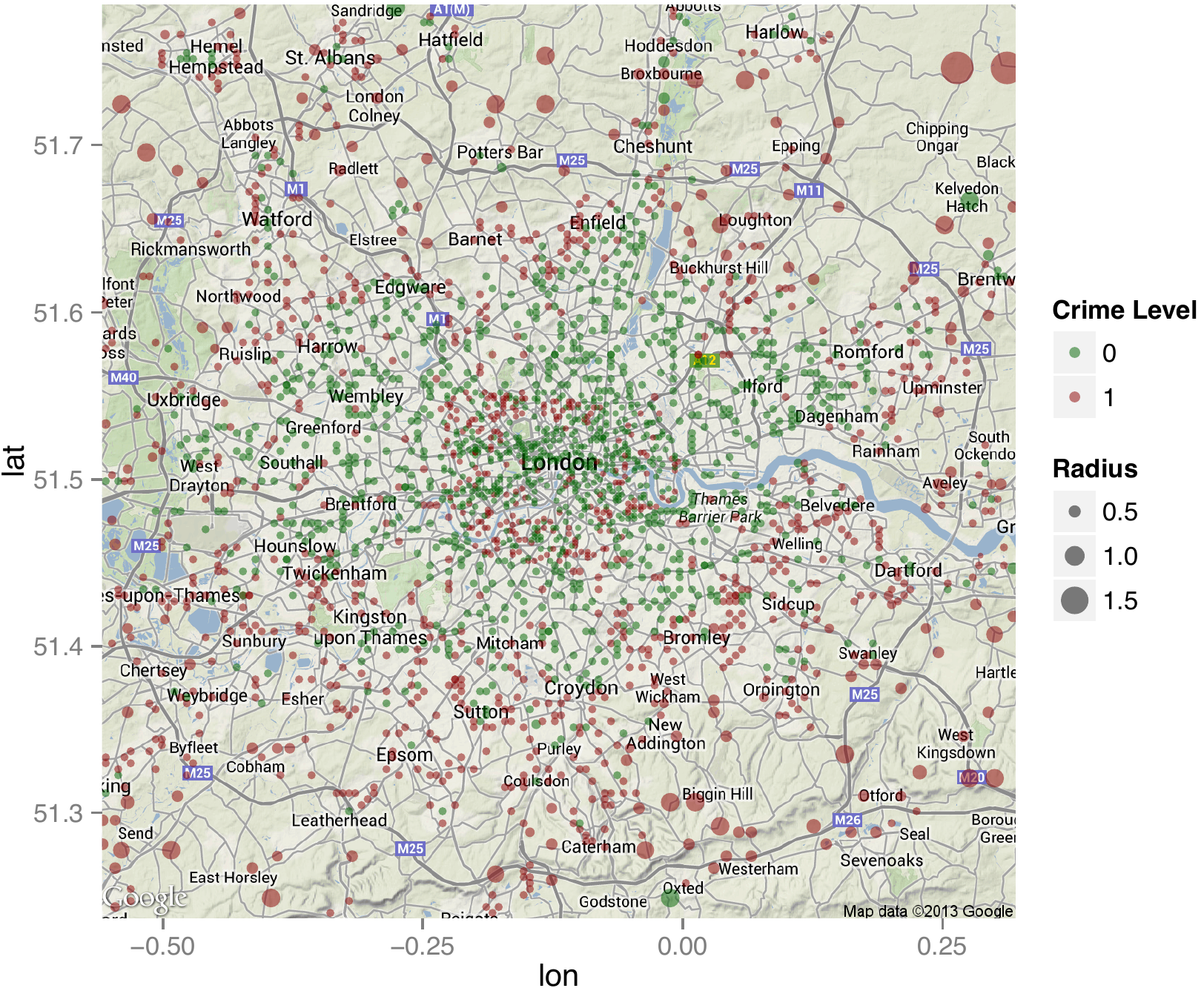}
\caption{Predicted Crime Hotspots}
\label{fig:crimePredicted}
\end{minipage}
\end{figure*}

A spatial visualisation of our results is reported on a map of the London metropolitan area in Figure
\ref{fig:crimePredicted} and compared with a similar visualisation of the ground truth labels in Figure \ref{fig:crimeGroundTruth}. In the maps, green represents "low crime level" and red "high crime level".

Second order features, which we introduced to capture intertemporal dependencies for our problem, not only made the feature space more compact, but also yielded a significant improvement in model performance metrics.

%Predicting next month crime based on aggregated mobile network activity of a
%previous month, including some additional features from borough profiles, showed
%the following nice results which are spatially visualized on the map  (Figure
%\ref{fig:crimePredicted}) for each test set ground truth labels, which are
%plotted side by side map for a better comparison and visual perception (Figure
%\ref{fig:crimeGroundTruth}).

%Twitter data was not included for the final model 538-dimentional feature space
%after the feature selection process.

%The best final binary classification model applied
%to the described feature space in Section \ref{sec:Methodology} showed the following performance metrics
%(Table \ref{tab:metrics}).

%\begin{table}[ht]
% \caption{Random Forest Performance Metrics}
%\caption{Performance metrics for the best model}
%\label{tab:metrics} 
%\centering
%\begin{tabular}{r|r}
%  \hline
%Accuracy & 0.6954 \\
%95\% CI & (0.6788, 0.7117) \\%
%P-Value $[Acc > NIR]$ & $<$2e-16 \\
%Cohen's Kappa & 0.3879 \\
% Mcnemar's Test P-Value & 0.3649 \\
%Area under the ROC curve & 0.6384 \\
%F1 & 67.2274 \\
%Sensitivity & 0.6669 \\
%Specificity & 0.7206 \\
% Pos Pred Value & 0.6714 \\
% Neg Pred Value & 0.7038 \\
% Prevalence & 0.4684 \\
% Detection Rate & 0.3081 \\
% Detection Prevalence & 0.4590 \\
%   \hline
%\end{tabular}
%\end{table}

 % %
% \begin{figure}[ht] \centering{
% \includegraphics[scale=0.8]{fig/mapGT.pdf}}
% \caption{Crime Ground Truth}
% \label{fig:crimeGroundTruth}
% \end{figure}
% %
% %
% \begin{figure}[ht] \centering{
% \includegraphics[scale=0.8]{fig/mapP.pdf}}
% \caption{Crime Predicted}
% \label{fig:crimePredicted}
% \end{figure}
% %
%

In order to understand the value added by the Smartsteps data, we compared the performance of the Random Forest
using all features with two different models trained with (i) only the subset of selected features
derived from the borough profiles dataset (Borough Profiles) and (ii) only the subset of selected features derived from the Smartsteps dataset (Smartsteps). 

Table \ref{tab:MetricsComparison} reports accuracy, F1, and the area under the ROC curve metrics for each of the models. In this Table, we also report the performance of (iii) a simple majority classifier, which always returns the majority class ("High Crime") as prediction (accuracy=53.15\%). 

%Finally, a visual comparison of the ROC curves for each of the models is provided in Figure \ref{fig:roc}. 

%\nuria{the table shows a small improvement of the smartsteps+borough model!!}
As can be seen on Table \ref{tab:MetricsComparison}, the borough-only model yields an accuracy of 62.18\%, over 6\% lower than the accuracy obtained with the Smartsteps model (68.37\%). 
%the addition of the borough profiles features does not add significant predictive power to the Smartsteps model. T
%As can be seen on the table, the Smartsteps model yields an increase in accuracy of almost 7.7\% when compared with the borough profiles model (69.9\% vs 62.2\% accuracy). 
The Smartsteps+Borough model yields an increase in accuracy of over 7\% when compared with the borough profiles model (69.54\% vs 62.18\% accuracy) while using the same number of variables. 

%We infer that ``Smartsteps + Borough Profiles'' Model improves prediction performance a lot and is feasible for crime prediction modelling tasks using aggregated mobile network data.

%\nuria{I already updated the table with the 68 feature models}
\begin{table}[H]
\caption{Metrics Comparison}
\label{tab:MetricsComparison}
\centering
\resizebox{\columnwidth}{!}{
%\scalebox{0.75}{
	%\begin{tabular}{l|l|r|r|r|r|r}	
	\begin{tabular}{l|r|r|r|r}
% 	  \hline
% 	 & Model & Accuracy, \% & F1, \% & Kappa, \% & AUC \\ 
% 	  \hline
%   1 & Baseline Equiprobable & 48.2410423452769 & 46.6241182398388 & -3.50087180203113 & 0.500583927301702 \\
%   2 & Baseline Empirical Probabilities & 50.4885993485342 & 47.4775397373877 & 0.654102382649463 & 0.501005803656014 \\ 
%   3 & Borough Profiles Model & 62.5732899022801 & 57.9582875960483 & 24.3996322245059 & 0.58080805780986 \\ 
%   4 & Smartsteps + Borough Profiles Model & 68.9250814332248 & 66.384778012685 & 37.5028274146121 & 0.633108445732036 \\ 
% 	   \hline
  \hline
 %& 
 Model & Acc.,\% & Acc. CI, 95\% & F1,\% & AUC \\ 
  \hline
  %1 & 
  %Baseline Equiprobable & 49.25 & (0.49, 0.52) & 47.44 & -1.54 & 0.51 \\ 
  %2 & 
  Baseline Majority Classifier & 53.15 & (0.53, 0.53) & 0 & 0.50 \\ 
  %3 & 
  Borough Profiles Model (BPM) & 62.18 & (0.61, 0.64) & 57.52 & 0.58 \\ 
%  4 & Smartsteps  & 69.9023 & (0.6825, 0.7152) & 67.6697 & 39.5182 & 0.6416 \\ 
%  5 & Smartsteps + Borough Profiles Model & 70.1954 & (0.6854, 0.7181) & 68.1518 40.1445 & 0.6442 \\ 
  %4 & 
  Smartsteps  & 68.37 & (0.67, 0.70) & 65.43 & 0.63 \\ 
  %5 & 
  Smartsteps + BPM & 69.54 & (0.68, 0.71) & 67.23 & 0.64 \\ 
   \hline
\end{tabular}
}
\end{table}

%\begin{figure}
%\label{fig:roc}
%\centering
% \includegraphics[scale=0.36]{fig/roc.pdf}
%\includegraphics[scale=0.36]{fig/roc_wo.pdf}
%\caption{Area under the ROC Curve}
%\end{figure}

\begin{table*}[!ht]
\caption{Top-20 Selected Features Ranked by Mean Decrease in Accuracy}
\label{tab:top30features}
\centering
\scalebox{0.75}{
\begin{tabular}{rl|rrrr}
  \hline
 Rank & Features & 0 & 1 & MeanDecreaseAccuracy & MeanDecreaseGini \\ 
  \hline
1 & smartSteps.daily.ageover60.entropy.empirical.entropy.empirical & 4.48 & 5.43 & 9.02 & 18.75 \\ 
  2 & smartSteps.daily.athome.mean.sd & 3.20 & 7.60 & 8.91 & 27.13 \\ 
  3 & smartSteps.daily.age020.sd.entropy.empirical & 5.69 & 3.97 & 8.85 & 16.88 \\ 
  4 & smartSteps.daily.age020.mean.entropy.empirical & 3.09 & 5.88 & 8.85 & 17.26 \\ 
  5 & smartSteps.daily.age020.mean.sd & 4.50 & 5.27 & 8.65 & 16.03 \\ 
  6 & smartSteps.daily.athome.min.entropy.empirical & 6.39 & 2.32 & 8.61 & 15.99 \\ 
  7 & smartSteps.daily.athome.sd.sd & 3.22 & 8.58 & 8.60 & 45.82 \\ 
  8 & smartSteps.daily.athome.sd.mean & 3.35 & 5.83 & 8.57 & 24.93 \\ 
  9 & smartSteps.daily.ageover60.entropy.empirical.sd & 4.62 & 4.95 & 8.56 & 20.45 \\ 
  10 & smartSteps.daily.athome.sd.median & 5.41 & 5.04 & 8.50 & 26.48 \\ 
  11 & smartSteps.daily.age3140.entropy.empirical.max & 2.33 & 5.79 & 8.44 & 16.24 \\ 
  12 & smartSteps.daily.age3140.min.sd & 6.81 & 4.06 & 8.31 & 36.52 \\ 
  13 & smartSteps.daily.athome.min.sd & 4.36 & 6.85 & 8.29 & 34.26 \\ 
  14 & smartSteps.daily.athome.sd.max & 4.13 & 6.87 & 8.27 & 34.89 \\ 
  15 & smartSteps.monthly.athome.max & 3.92 & 5.42 & 8.26 & 29.86 \\ 
  16 & smartSteps.monthly.athome.sd & 4.43 & 4.17 & 8.21 & 39.70 \\ 
  17 & smartSteps.daily.age5160.entropy.empirical.entropy.empirical & 4.74 & 4.11 & 8.13 & 16.64 \\ 
  18 & smartSteps.daily.age020.sd.sd & 3.67 & 5.88 & 8.12 & 16.86 \\ 
  19 & smartSteps.daily.athome.entropy.empirical.entropy.empirical & 5.13 & 4.82 & 8.08 & 18.55 \\ 
  20 & smartSteps.daily.athome.max.sd & 2.83 & 6.29 & 8.07 & 26.85 \\
   \hline
\end{tabular}
}
\end{table*}

\section{Discussion}
\label{sec:Discussion}

%\bruno{the discussion of the results is really complex ... we can only state that entropy is useful for predicting crime level but we cannot say anything about the direction of this association ... for example low or high entropy is associated to low crime? My suggestion is to report some correlations among crime level and the more predictive features ... I know that we worked in a non-linear space for our prediction but without this kind of information there is almost nothing to discuss}
The results discussed in the previous section show that usage of human behavioral data (at a daily and monthly scale) significantly improves prediction accuracy when compared to using rich statistical data about a borough's population (households census, demographics, migrant population, ethnicity, language, employment, etc...). The borough profiles 
data provides a fairly detailed view of the 
nature and living conditions of a particular area in a city, yet it is expensive and effort-consuming to collect. 
Hence, this type of data is typically updated with low frequency (\emph{e.g.} every few years). Human behavioral data derived from mobile network activity and demographics, though less comprehensive than borough profiles, provides significantly finer temporal and spatial resolution. 

Next, we focus on the most relevant predictors of crime level which show interesting associations. We first take a look at the top-20 variables in our model, which are sorted by their mean reduction in accuracy (see Table \ref{tab:top30features}). % for reader convenience.

The naming convention that we used for the features shown in Table \ref{tab:top30features} is: the original data source (\emph{e.g.} "smartSteps") is followed by the temporal granularity $T$ (\emph{e.g.} "daily"), the  semantics of the variable (\emph{e.g.} "athome"), and its statistics at $T$ (\emph{e.g.} "mean"). Note that \emph{second order} features where we computed statistics across multiple days appear \emph{after} 
the first statistics. For example, feature 2 in the Table, "smartSteps.daily.athome.mean.sd", is generated by computing the standard deviation of the daily means of the percentage of people estimated to be at home. 
%the std of the daily means of people estimated to be at home
%variables extracted at a daily  level a further field is present that encodes statistical information of the latter over the sequence of days we take into account.
%\nuria{explain notation and meaning of variables}
%\jacopo{tried.. see above}

As shown in the Table, the Smartsteps features have more predictive power than official statistics coming from borough profiles. No features listed in the top-20 are actually obtained using borough profiles. Moreover, Table
\ref{tab:top30features} shows that higher-level features extracted 
%on a daily basis
over a sequence of days from variables encoding the daily dynamics 
(all features with the label smartSteps.daily.*) 
%\nuria{please add one feature as an example}
have more predictive power than features extracted on a monthly basis.
For example, feature 2 in Table~\ref{tab:top30features}. 
This finding points out at the importance of capturing the temporal dynamics of a geographical area in order to predict its levels of crime.
%\nuria{this is kind of interesting and strange given that the crime data is only provided at a monthly level}\jacopo{fixing: in fact these higher-lever features describe the dynamics over consecutive days of features computed on a daily basis. tried to put it in a decent way. still not nice to read. }

Furthermore, features derived from the percentage of people in a certain 
cell who are at home (all features with \emph{.athome.} in their label) 
both at a daily and monthly basis seem to be of extreme importance. 
In fact, 11 of the top 20 features are related to the \emph{at home} variable. 
%\nuria{we should probably connect this with previous work on crime modeling}

It is also interesting to note the role played by diversity patterns captured by Shannon entropy features \cite{shannon48}. The entropy-based features (all features with \emph{.entropy.} in their label) in fact seem useful for predicting the crime level of places (8 features out of the top 20 are entropy-based features). In our study, the Shannon entropy captures the predictable structure of a place in terms of the types of people that are in that area over the course of a day. A place with high entropy would have a lot of variety in the types of people visiting it on a daily basis, whereas a place with low entropy would be characterised by regular patterns over time. In this case, the daily diversity in patterns related to different age groups, different use (home vs work) and different genders seems a good predictor for the crime level in a given area. Interestingly, Eagle \emph{et al.} \cite{Eagle2010} found that Shannon entropy used to capture the social and spatial diversity of communication ties within 
an individual's social network was strongly and positively correlated with economic development. 

%Looking in some more detail, we can see that some of the features have a strong predictive power only to predict one of the two classes (see columns labeled '0' and '1' in Table \ref{tab:top30features}). For instance, the entropy in the percentage of females inside a cell of the grid (feature 22 on the table) has strong discriminative power when predicting high crime levels whereas its discriminative power to predict low crime levels is not significant. Similarly, entropy features for the 31-40 and 51-60 age brackets (features 25 and 18) play a stronger predictive role for high crime areas than for low crime areas. Conversely, entropy features for the 0-20 age bracket (feature 16) play a stronger predictive power for low crime areas than for high crime areas.

%\bruno{Andrey/All, for me it's really complex understand the meaning of some of these complex features and so use them to extract relevant indicators for high and low crime areas}

As previously described, borough profile features (official statistics) have lower predictive 
power with respect to accuracy than features extracted from 
aggregated mobile network activity data. Six borough profile features were selected in the final feature vector, including the proportion of the working age population who claim out of work benefits, the largest migrant population, 
the proportion of overseas nationals entering the UK and 
the proportion of resident population born abroad --metrics based on 2011 Census Bureau data. The predictive power of some of these variables is in line with previous studies in sociology and criminology. For example, several studies show a positive association among unemployment rate and crime level of an area \cite{Raphael2001}. Still under debate is the positive association among number of immigrants and crime level \cite{Ellis2009}.
However, our experimental results show that the static nature of these variables makes them less useful in predicting crime level's of a given area when compared with less detailed but daily information about the types of people present in a same area throughout the day. 

%Finally, it is to note that Twitter features had very low predictive power when compared to the other features: as previously described, no Twitter-derived features were selected to be included in our final prediction model. This is probably due to the coarse spatial granularity of the Twitter data (neighborhood level versus Smartsteps cell) and the small sample of tweets that was available.
%as part of the Datathon. 

\section{Implications and Limitations}
We have outlined and tested a multimodal approach to automatically predict with almost 70\% accuracy whether a given geographical area will have high or low crime levels in the next month. The proposed approach 
could have clear practical implications by informing police departments and city governments on how and where to invest their efforts and on how to react to criminal events with quicker response times. From a proactive perspective, the ability to predict the safety of a geographical area may provide information on explanatory variables that can be used to identify underlying causes of these crime occurrence areas and hence enable officers to intervene in very narrowly defined geographic areas.

The distinctive characteristic of our approach lies in the use of features computed from aggregated and anonymized mobile network activity data in combination with some demographic information. Previous research efforts in criminology have tackled similar problems using background historical knowledge about crime events in specific areas, criminals' profiling, or wide description of areas using socio-economic and demographic indicators. Our findings provide evidence that aggregated and anonymized data collected by the mobile infrastructure contains relevant information to describe a geographical area in order to predict its crime level.

The first advantage of our approach is its predictive ability. Our method predicts crime level using variables that capture the dynamics and characteristics of the demographics and nature of a place rather than only making extrapolations from previous crime histories. Operationally, this means that the proposed model could be used to predict new crime occurrence areas that are of similar nature to other well known occurrence areas. Even though the newly predicted areas may not have seen recent crimes, if they are similar enough to prior ones, they could be considered to be high-risk areas to monitor closely. This is an important advantage given that in some areas people are less inclined to report crimes \cite{Tarling2010}.
Moreover, our approach provides new ways of describing geographical areas. Recently, some criminologists have started to use \emph{risk terrain modeling} \cite{Caplan2010} to identify geographic features that contribute to crime risk, \emph{e.g.} the presence of liquor stores, certain types of major stores, bars, etc. Our approach can identify novel risk-inducing or risk-reducing features of geographical areas. In particular, the features used in our approach are dynamic and related to human activities. 

Our study has several limitations due to the constraints of the datasets used. 
%released during the Datathon for Social Good.
First of all, we had access only to 3 weeks of Smartsteps data collected between December 2012 and the first week of January 2013. In addition, the crime data provided was aggregated on a monthly basis. Having access to crime events aggregated on a weekly, daily or hourly basis would enable us to validate our approach with
finer times granularity, predicting crimes in the next week, day or even hour.

% \section{Discussion}
% \label{sec:Discussion}
% 
% {\ldots}
% 
\section{Conclusion}
\label{sec:Conclusion}
In this paper we have proposed a novel approach to predict crime hotspots from human behavioral data derived from mobile network activity, in combination with demographic information. Specifically, we have described a methodology to automatically predict with almost 70\% of accuracy whether a given geographical area of a large European metropolis will have high or low crime levels in the next month. We have shown that our approach, while using a similar number of variables, significantly improves prediction accuracy (6\%) when compared with using traditional, rich --yet expensive to collect-- statistical data about a borough's population. 
Moreover, we have provided insights about the most predictive features (\emph{e.g.} home-based and entropy-based features) and we have discussed the theoretical and practical implications of our methodology.
Despite the limitations discussed above and the additional investigations needed to validate our approach and the robustness of our indicators, we believe that our findings open the door to exciting avenues of research in computational approaches to deal with a well-known social problem such as crime.

% ACKNOWLEDGMENTS are optional
\section{Acknowledgments}
\small
Source data processing, reported data transformations and feature selection, which were derived from anonymised and aggregated mobile network dataset, provided by Telef\'{o}nica Digital Limited
%, a corporation existing under the laws of England, with company
%number 07884976 and a registered office or regular place of business located at
%260 Bath Road, Slough, Berkshire, SL1 4DX, 
were done only during the public
competition -- ``Datathon for Social Good'' organized by Telef\'{o}nica, The Open Data Institute and the MIT during the Campus Party Europe 2013 at the O2 Arena in London during 
2-7 September 2013, in full compliance with the competition rules and legal limitations imposed by the ``Terms and Conditions''
document. 
%Data analysis of open datasets was also done during this paper
%preparation timeframe.

The work of Andrey Bogomolov is partially supported by EIT ICT
Labs Doctoral School grant and Telecom Italia Semantics
and Knowledge Innovation Laboratory (SKIL) research grant T.

\end{document}